\documentclass[%
 reprint,
preprintnumbers,
 amsmath,amssymb,
 aps,
]{revtex4-2}

\usepackage{graphicx}% Include figure files
\usepackage{dcolumn}% Align table columns on decimal point
\usepackage{bm}% bold math
\usepackage[hidelinks]{hyperref}% add hypertext capabilities
\usepackage{amsmath}
\usepackage{xcolor}
\DeclareUnicodeCharacter{2212}{-}

\begin{document}

\preprint{ADP-22-25/T1196}
\title{Dense Nuclear Matter with Baryon Overlap}%
 
\author{Jesper Leong}
\author{Theo F. Motta}
\altaffiliation[Now at the ]{Institut f\"{u}r Theoretische Physik, Justus-Liebig-Universit\"{a}t Giessen. }%\\ Humboldt Foundation}%Lines break automatically or can be forced with \\
\author{Anthony W. Thomas}
 \affiliation{%
 CSSM and ARC Centre of Excellence for Dark Matter Particle Physics, Department of Physics, University of Adelaide, SA 5005 Australia}%
\author{P.A.M. Guichon}
 \affiliation{{\it Irfu}, CEA, Universit\'e Paris-Saclay, F91191 Gif sur Yvette France}

\date{\today}% It is always \today, today,

\begin{abstract}
The possibility of new short-distance physics applicable inside the cores of NS is incorporated into the equation of state generated by the quark-meson coupling model. The contribution of this new physics to the energy density is taken to be proportional to the amount of overlap between the quark cores of the baryons involved. With no change to the properties of symmetric nuclear matter at saturation density, including an incompressibility compatible with data on giant monopole resonances, one can sustain neutron stars with a maximum mass $M_{max}>2.1$ M$_\odot$, even when hyperons are included. 
\end{abstract}

\keywords{neutron stars; equation of state of dense matter; hyperons; short-distance repulsion; baryon overlap }%Use showkeys class option if keyword
\maketitle

\section{Introduction}
Understanding the equation of state (EoS) of dense nuclear matter, namely the pressure as a function of energy density~\cite{Glendenning:1997wn,Shapiro:1983du,Stone:2021uju}, $p=p(\epsilon)$, is one of the key challenges being addressed in modern nuclear theory. The EoS is well known in the low density
region~\cite{Gandolfi:2009fj,Kruger:2013kua} and is believed to be constrained by perturbative QCD in the high density region~\cite{Annala:2019puf,Kojo:2014rca,Kojo:2015nzn,Kurkela:2014vha}. In the intermediate region, the description of nuclear matter is still a matter of considerable debate. Heavy ion reactions have provided important constraints \cite{Danielewicz:2002pu,Stone:2022mho} up to densities of order several times the saturation density of symmetric nuclear matter, $n_0$. However, the matter formed in these collisions exists only for a short time and in a small volume, which introduces model dependence in the interpretation. Intense interest has shifted to long lived, cold neutron stars (NS) because their core densities are expected to be as large as 4-10 times $n_0$. At such large densities we cannot be sure whether the matter is hadronic, quark matter or some hybrid form \cite{Masuda:2012kf,Motta:2022nlj,Annala:2019puf}. Furthermore, as the matter in a NS is stable long-term, to the extent that it is hadronic, one must satisfy the conditions of $\beta$-equilibrium and hyperons must be present.

Of the more than 40 NS mass measurements, most centre around the canonical mass of $1.4-1.5$ M$_\odot$ \cite{Freire,Ozel:2016oaf,Motta:2022nlj,Steiner:2014pda}. These rely on the detection of a NS in a binary system~\cite{Ozel:2016oaf,Chatterjee:2015pua,Burgio:2021vgk}. Given the compactness and the extraordinary distances of these stars from Earth, mass and radius measurements on the same star still carry considerable uncertainty, making them sub-optimal for constraining the EoS. Much of the focus has shifted to the heaviest of these stars, of which 3 are known; PSR J0348+0432 has a mass  $M=2.01^{+0.04}_{-0.04}$ M$_\odot$~\cite{Antoniadis:2013pzd}, PSR J1614-2230 has $M=1.908^{+0.016}_{-0.016}$ M$_\odot$~\cite{NANOGrav:2017wvv} and PSR J0740+6620 has  $M=2.072^{+0.067}_{-0.066}$ M$_\odot$~\cite{Riley:2021pdl,Fonseca:2021wxt}.

Measurements of NS radii are particularly difficulty to narrow down~\cite{Al-Mamun:2020vzu}. However, the gravitational wave (GW) detection of GW170817 has placed new emphasis on the radii of NS~\cite{LIGOScientific:2018cki}. The NS merger event also allowed a totally new property, the tidal deformability, to be extracted from the waveform of the passing gravitational wave. For a $1.1-1.5$ M$_\odot$ NS, the radius is largely independent of the mass. GW170817 involved component masses of $m_1\in [1.15, 1.46]$ M$_\odot$ and $m_2\in [1.36, 1.62]$ M$_\odot$, with radii for either star $R=11.9\pm1.4$ km, making it highly likely that the event involved a binary NS system~\cite{LIGOScientific:2018cki}. (GW190425 may contain NS but there is a high probability that the merger event has at least one black hole component~\cite{LIGOScientific:2020aai}.) The tidal deformability measured for GW170817, which was reported to be $\Lambda_{1.4}=190^{+390}_{-120}$, is a measure of the quadrupole deformation of a spherical object caused by an inhomogenous external gravitational
field~\cite{LIGOScientific:2018cki}. GW measurements have the further advantage that they may offer insights concerning the presence of exotic matter in 
stars~\cite{Chatziioannou:2015uea,Chatziioannou:2020pqz}.

With the observation of NS with masses $M\approx2$ M$_\odot$ many EoS have been ruled out \cite{Antoniadis:2013pzd,Demorest:2010bx,Riley:2021pdl}. This is particularly true for those where hyperons are included. Under $\beta$-equilibrium, hyperons are predicted to appear at or above 3 $n_0$, and because they have low momentum near the threshold for their introduction, the pressure corresponding to a given energy density is lower than it would be without hyperons. Then the maximum mass of the star must be lower~\cite{Motta:2022nlj,Chatterjee:2015pua,Burgio:2021vgk,Baldo:2011gz}. 

The quark-meson-coupling (QMC) model, is unique amongst models for the binding of nuclear matter in that the change in the structure of any bound hadron, induced by the strong mean scalar field in-medium~\cite{Guichon:1987jp,Guichon:1995ue}, plays a key role. 
In particular, the resultant density dependent reduction in the coupling of baryons to the scalar field~\cite{Guichon:2006} is equivalent to introducing many-body forces~\cite{Guichon:2004xg}. The three-body force, in particular, acts between all combinations of baryons, NNN, YNN, YYN and YYY (with Y a hyperon), without the introduction of any new parameters. As a result of this three-body force, the QMC model predicted the existence of NS with masses up to $2$ M$_\odot$, with hyperons~\cite{Stone2007}, three years before the first was observed~\cite{Demorest:2010bx}.

Following the discovery of NS with masses as large as $2$ M$_\odot$, many possible solutions to what was regarded as the ``hyperon puzzle'' have been proposed. These include density dependent couplings, many body forces and a possible phase transition from hadronic to deconfined quark matter \cite{Glendenning:1992vb,Whittenbury:2015ziz}. By interpolating between accurate low ($n<1.1$ $n_0$) and high ($n>40$ $n_0$) density EoS, several researchers have derived phenomenological EoS matching heavy NS observations~\cite{Annala:2019puf,Kojo:2014rca,Kojo:2015nzn,Kurkela:2014vha}. The requirement that the EoS approaches that of perturbative QCD at high densities has a distinct imprint on the speed of sound ($c_s$). Conformal matter has a limit of $c_s^2<1/3$ approaching from below, which suggests deconfined quark matter~\cite{Annala:2019puf}, although the appearance of hyperons has a similar signature~\cite{Motta:2020xsg,Stone:2019blq}.

The QMC model has been extensively developed since the prediction of high mass NS with hyperons. The introduction of an isovector scalar meson did not dramatically change the predictions for NS properties, with the most significant change being an increase in the NS radius, for a given mass~\cite{Motta:2019tjc}. 

The energy density functional derived in the model~\cite{Guichon:2006,Guichon:2018uew} has been applied to the properties of even-even nuclei across the periodic table, with considerable success~\cite{Martinez:2018xep,Martinez:2020ctv,Stone:2019syx,Stone:2017oqt}. However, studies of the giant monopole resonances required the introduction of a term cubic in the scalar potential, which also lowered the incompressibility of nuclear matter. The same term included in calculations of the EoS of dense nuclear matter tends to lower the maximum neutron star mass.

The parameters of the QMC model, associated with the exchanged mesons, are determined by the properties of infinite nuclear matter at saturation density~\cite{Stone:2022unw}. On the other hand, the cores of the highest mass NS involve much higher densities, where the finite-size baryons may overlap; something not included in the model. In that region non-perturbative gluon exchanges may lead to additional repulsion not accounted for by longer-range meson exchange. With this in mind, here we extend the QMC model to include a new, phenomenological, short-ranged repulsive force, which only affects the EoS at densities well above saturation density. 

The structure of this paper is as follows. In section~\ref{sec:theory} we summarise the calculation of the EoS of dense matter within the QMC model and introduce the new short-distance repulsive force, designed to affect the EoS only at densities well above nuclear matter density. The results and concluding remarks are presented in sections~\ref{sec:Results} and \ref{sec:Conclusion}.
 
%%%%%%%%%%%%%%%%%%%%%%%%%%%%%%%%%%%%%%%%%%%
%%%%%%%  QMC Review %%%%%%%%%%%%
\section{Theoretical framework}
\label{sec:theory}
In the MIT bag model, the quarks are confined in a spherical region, the interior of which is a strictly perturbative vacuum. In the QMC model the baryons retain their individual identity, even at higher 
densities~\cite{Chodos:1974pn,Thomas:1982kv}. One of the criticisms which Stone~\textit{et al.}~\cite{Stone2007} highlighted is the possibility that the QMC model would begin to breakdown at some unknown threshold density, where the hadrons start to overlap and eventually transition into quark matter. Actually, this may even be apparent at lower densities, of order 4 $n_0$, where nucleons geometrically touch~\cite{Ozel:2016oaf}. Performing QCD lattice calculation, Bissey \textit{et al.} showed that three quarks in a baryon are connected by a Y-shaped flux tube~\cite{Bissey:2006bz}. Elsewhere a non-perturbative QCD vacuum exists, through which baryons are permitted to move freely. Thus a literal interpretation of the MIT bag and QMC models may not reflect the full physical picture of nuclear matter~\cite{Stone2007,Guichon:2018uew}. While this provides some justification for treating the baryons as non-overlapping to higher densities than one might naively expect, one may anticipate that new interactions are likely to be needed eventually.

Berryman and Gardner proposed an extension to the Standard Model based upon gauging baryon number~\cite{Berryman:2021jjt}. Their new short ranged, repulsive quark-quark force was mediated by a vector boson coupling to baryon number $B$. Normally this would be suppressed by the highly repulsive short range ($0.4-0.6$ fm) NN, YN and YY potential~\cite{Inoue:2010hs}. However, beyond saturation density, as the baryons start to \textit{overlap}, the new interaction would have an affect on the resulting EoS~\cite{Berryman:2021jjt,Berryman:2022zic}. To model this, a two particle NN Yukawa potential was used and Berryman and Gardner showed that the ensuing NS had a maximum mass of $2-2.2$ M$_\odot$~\cite{Berryman:2021jjt}.

In the QMC model the $\omega$ meson generates the short-distance repulsion between baryons. However, as explained earlier, as the baryons begin to overlap we expect that there may be additional repulsion arising from non-perturbative QCD in the multi-quark environment. To model this, we propose an additional term to be added to the QMC energy density, which only has appreciable effects at supra-nuclear densities. The functional form used to model the degree of overlap is taken to have a simple Gaussian form, given in Eq.~(\ref{eq:overlap}), and is required to stiffen the EoS without changing the physics at saturation density.
%Physically, the overlap term models a mixed phase between hadronic and deconfined quark matter. 

Our summary of the key details of the QMC model follows closely the work of Motta~\textit{et al.}~\cite{Motta:2019tjc} and Guichon~\textit{et al.}~\cite{Guichon:2018uew}. The calculation is carried out in the Hartree-Fock approximation, with the inclusion of the isovector scalar meson, $\delta$, as introduced by Motta 
\textit{et al.}~\cite{Motta:2019tjc}.

\subsection{Energy Density}
In infinite nuclear matter, using mean-field approximation, the baryon effective mass is given by 
\begin{equation}
    M^*_B=\frac{\Omega_uN_u+\Omega_dN_d+\Omega_sN_s-z_0}{R_B}+BV_B+\Delta E_M \, ,
    \label{eq:Meff}
\end{equation}
while the energy density associated with the strong interaction is
\begin{equation}
    \label{eq:baryonenergy3}
    E_{total}=\sqrt{M_B^{*2}(\bar{\sigma},\bar{\delta})+\vec{k}^2}
    +3 g^q_\omega \bar{\omega} + g^q_\rho I_B \bar{\rho} \, .
\end{equation}
Here, the mean iso-scalar and iso-vector scalar mean-fields are  $\bar{\sigma}$ and $\bar{\delta}$, respectively, while $\bar{\omega}$ and $\bar{\rho}$ are the time components of the iso-scalar and iso-vector vector mean-fields. $\Omega_i$ expresses the lowest eigenvalue for the three valence quarks, calculated in the local scalar mean-fields. The input parameters $B$, $\Delta E_M$, $z_0$ are the bag pressure, hyper-fine colour interaction and the zero point gluon fluctuation terms, fitted to satisfy the mass of the free baryon, while satisfying the boundary conditions imposed by the MIT bag model~\cite{Chodos:1974pn,Thomas:1984dp}. The effective mass, $M_B^*$ has previously been shown to be fit well up to quadratic order~\cite{Motta:2019tjc}
\begin{eqnarray}
M^*_B(\bar{\sigma},\bar{\delta})=&&M_B-w^\sigma_Bg_\sigma \bar{\sigma} +\tilde{w}^\sigma_B\frac{d}{2}(g_\sigma \bar{\sigma})^2 \nonumber \\
&&-t^\delta_Bg_\delta I_B \bar{\delta}+\tilde{t}^\delta_B(g_\delta \bar{\delta})^2 \nonumber \\
&&+\tilde{d}g_\sigma g_\delta \bar{\sigma} I_B \bar{\delta} \nonumber\\
    =&& M_B-g_\sigma^B(\bar{\sigma},\bar{\delta})\sigma-g_\delta^B(\bar{\sigma},\bar{\delta}) I_B \bar{\delta} \, .
    \label{eq:effM}
\end{eqnarray}

The scalar polarisabilities, $d$, $\tilde{d}$, and $\tilde{t}$, which are calculated using the underlying bag model and involve no new parameters, reflect the self-consistent response of the internal structure of the bound baryons to the applied scalar fields. They are at the origin of the repulsive three-body forces which arise naturally within the model~\cite{Guichon:2004xg,Guichon:2018uew}. The scalar fields change the effective mass and the vector fields shift the energy of the baryon. In the mean field approximation each of the meson field operators is set to its expectation value, while the corrections are treated as a small perturbation, $\sigma \rightarrow \langle \bar{\sigma}\rangle + \Delta\sigma$. Omitting the arguments in Eq.~(\ref{eq:effM}), the meson fields are then
\begin{eqnarray}
    %sigma
    m_{\sigma}^2 \bar{\sigma}
    &&= - 2\sum_B \frac{\partial{M^*_B}}{\partial\bar{\sigma}}
    \int^{k_f(B)} \frac{d^3k}{(2\pi)^3}\frac{M^*_B}{\sqrt{k^2+M^{*2}_B}} \\
    %omega
    m_\omega^2\bar{\omega}&&= \sum_B g^B_\omega n_B \\
    %rho
    m_\rho^2\bar{\rho} &&= \sum_B g_\rho^B I_B n_B \\ 
    %delta
     m_{\delta}^2 \bar{\delta}
    &&= - 2\sum_B \frac{\partial{M^*_B}}{\partial\bar{\delta}}
    \int^{k_f(B)} \frac{d^3k}{(2\pi)^3}\frac{M^*_B}{\sqrt{k^2+M^{*2}_B}}
    \, ,
\end{eqnarray}
where the summation runs over the baryon species ($n, p, \Lambda, \Sigma^{0,\pm}$). The integrals range from 0 to the relevant Fermi momentum, $k_f(B)$
\begin{eqnarray}
    \label{eq:Fermienergy}
    k_f(B)=\sqrt[3]{3\pi^2n_B} \, ,
\end{eqnarray}
with $n_B$ the number density. $I_B$ denotes the isospin of baryon $B$.

The contribution to the energy density from baryons and the meson fields, $\epsilon_{B}$, is:
\begin{eqnarray}
\label{eq:Edensity}
    \epsilon_B=\frac{\langle \mathcal{H}_B+\mathcal{V}_\sigma+\mathcal{V}_\omega+\mathcal{V}_\rho+\mathcal{V}_\delta+\mathcal{V}_\pi \rangle}{V}.
\end{eqnarray}
In the Hartree-Fock approximation, these are evaluated as follows
\begin{eqnarray}
    \label{eq:energyfreebaryon}
    \frac{\langle \mathcal{H}_B \rangle}{V} = &&
    2 \sum_B \int^{k_F} \frac{d^3k}{(2\pi)^3} \sqrt{{\vec{k}}^2+{M^*_B}^2}
    \\
    \label{eq:Hsigma}
    \frac{\langle \mathcal{V}_\sigma \rangle}{V}=&&
    \frac{{m_\sigma}^2\bar{\sigma}^2}{2} + \frac{\lambda_3}{3!}g_\sigma^3\bar{\sigma}^3 + \sum_B \left( \frac{\partial M^*}{\partial\bar{\sigma}} \right) ^2 \nonumber \\
    &&\times\int^{k_F} \frac{d^3k_1d^3k_2}{(2\pi)^6} \frac{1}{(\vec{k_1}-\vec{k_2})^2+m_\sigma^2} \frac{M^*_BM_{B'}^*}{E_{B}E_{B'}} \\
    \label{eq:Homega}
    \frac{\langle \mathcal{V}_\omega \rangle}{V} =&& 
    \frac{m_\omega^2\bar{\omega}^2}{2} -\sum_B g_{\omega}^2\int^{k_F}\frac{d^3k_1d^3k_2}{(2\pi)^6}\frac{1}{(\vec{k_1}-\vec{k_2})^2+m_\omega^2}\nonumber \\
    && \\
    \label{eq:Hrho}
    \frac{\langle \mathcal{V}_\rho \rangle}{V}=&&
    \frac{m_\rho^2\bar{\rho}^2}{2} \nonumber \\ &&-\sum_B\sum_{i,j}{g_\rho}^2C_{i,j}\int^{k_F}\frac{d^3k_1d^3k_2}{(2\pi)^6}\frac{1}{(\vec{k_1}-\vec{k_2})^2+m_\rho^2} \nonumber \\ &&
    \\
    \label{eq:Hdelta}
    \frac{\langle \mathcal{V}_\delta \rangle}{V} =&& \frac{m^2_\delta \bar{\delta}^2}{2} + \nonumber \\
    &&
    \sum_B \int^{k_F} \frac{d^3k_1 d^3k_2}{(2\pi)^6} \frac{Z_{i,j}}{(\vec{k_1}-\vec{k_2})^2+m^2_\sigma} \frac{(M^*_BM_{B'}^*)}{E_B E_{B'}} \, ,  \nonumber \\
    &&
\end{eqnarray}
with $E_B$ the relativistic energy of baryon $B$.
Equation~(\ref{eq:Hsigma}) includes the term cubic in the $\sigma$ field required to lower the incompressibility of symmetric nuclear matter sufficiently to yield acceptable giant monopole resonance energies. The solutions of Eqs.~(\ref{eq:Homega}) and (\ref{eq:Hrho}) are simple to evaluate, because they correspond to 2-body interactions. Eqs.~(\ref{eq:Hsigma}) and (\ref{eq:Hdelta}) are more complicated, because the $\sigma$ and $\delta$ mean-fields are functions of $M^*_B$, which in turn depend on $\bar{\sigma}$ and $\bar{\delta}$. These are solved self-consistently at $\beta$-equilibrium for each given number density (see Stone \textit{et al.}~\cite{Stone2007}). The notation for $C_{i,j}$ and $Z_{i,j}$ is given in terms of the Kronecker delta ($\delta_{i,j}$) and the isospin eigenvalues ($I_B$) of the baryon, consistent with Ref.~\cite{Motta:2019tjc}.
\begin{eqnarray}
    C_{i,j}&&= \delta_{i,j}I_B^2+I_B(\delta_{i,j+1}+\delta_{i+1,j}) \\
    Z_{i,j}&&= \delta_{i,j}\frac{\partial M^*_i}{\partial \delta}\frac{\partial M^*_j}{\partial \delta}\nonumber \\
    &&+\left(\delta_{i,j+1}+\delta_{i+1,j}\right)g^{B_i}_\delta(\bar{\sigma},\bar{\delta})g^{B_j}_\delta(\bar{\sigma},\bar{\delta}) \, .
    \label{eq:Zij}
\end{eqnarray}

Finally the long range pion Fock terms~\cite{Stone2007,Krein:1998vc} are 
\begin{eqnarray}
    \frac{\left\langle \mathcal{V}_{\pi}\right\rangle}{V}=&&\frac{1}{n_{B}}\left(\frac{g_{A}}{2 f_{\pi}}\right)^{2}\left\{ \frac{}{} J_{p p}+4 J_{p n}+J_{n n} \right.
    \nonumber \\
%    &&-\frac{24}{25}\left(J_{\Lambda, \Sigma^{-}}+J_{\Lambda, \Sigma^{0}}+J_{\Lambda, \Sigma^{+}}\right)  \nonumber \\ 
%    &&+\frac{16}{25}\left(J_{\Sigma^-\Sigma^{-}}+2 J_{\Sigma^-\Sigma^{0}}+2 J_{\Sigma^{+} \Sigma^{0}}+J_{\Sigma^{+} \Sigma^{+}}\right)  \nonumber \\
    &&\left.+\frac{1}{25}\left(J_{\Xi^-\Xi^{-}}+4 J_{\Xi^-\Xi^{0}}+J_{\Xi^{0} \Xi^{0}}\right)\right\},
\end{eqnarray}
where
\begin{eqnarray}
    J_{f f^{\prime}}=\frac{1}{(2 \pi)^{6}} \int^{k_{F}(f) k_{F}\left(f^{\prime}\right)} d \vec{k} d \vec{k}^{\prime}\left[1-\frac{m_{\pi}^{2}}{\left(\vec{k}-\vec{k}^{\prime}\right)^{2}+m_{\pi}^{2}}\right] \nonumber \\
\end{eqnarray}
and the $\delta$-function associated with the first term in the square brackets is dropped because of short-distance repulsion (so-called ``poor man's absorption'').

The bag overlap term, denoted $\mathcal{H}_O$, is approximated by the overlap of simple Gaussian wave functions, with the overall strength, $E_0$, and range, $b$, treated as free parameters, subject to the constraint that this term must not change nuclear matter properties at saturation density. 
\begin{eqnarray}
\label{eq:overlap}
    \frac{\langle \mathcal{H}_O \rangle}{V}=E_0n_B^{}\text{exp}\left\{- \left(\frac{n_B^{-1/3}}{b}\right)^2\right\} \, .
\end{eqnarray}
We note that $n_B^{-1/3}$ is roughly the average distance between baryons in a Fermi gas.
Given that $b$ characterises the size of the quark cores of the baryons we expect it to be of order 0.5 fm.
This additional term is added to Eq.~(\ref{eq:Edensity}). The overlap term is assumed independent of the quark content of the baryon and is repulsive. This concludes the construction of the energy density for nuclear matter within the QMC framework.

\subsection{\texorpdfstring{$\beta$-Equilibrium}{beta-equilibrium}}
The material in the NS is assumed to be cold and in $\beta$-equilibrium.  Any particulate species which exist longer than the time scale of the system then participate to minimise the energy density. This includes the hyperons which are stable at high densities, because of Pauli-blocking. The total energy density of the system, including the leptons (electron and muon), is minimised under the condition of $\beta$-equilibrium. Thus the total energy density is
\begin{equation}
    \epsilon_{total}= \epsilon_B(n_B)+\epsilon_e(n_e)+\epsilon_\mu(n_\mu) 
    \, ,
\end{equation}
where $\epsilon_B$ is given by Eqs.~(\ref{eq:Edensity}) and (\ref{eq:overlap}). The energy density of the electrons and muons is described by a free gas of leptons
\begin{eqnarray}
\epsilon_l(n_l)=2\int^{k_l(l)}\frac{d^3k}{(2\pi)^3}\sqrt{\vec{k}^2+m_l^2}
\, .
\end{eqnarray}

Charge neutrality is imposed along with conservation of baryon number. As previous studies have shown that the $\Sigma^{0,\pm}$ baryons are absent for $n_B<1.2$ fm$^{-3}$~\cite{Stone2007,Whittenbury:2013wma}. The $\Sigma$ baryons have been removed from the calculation. The equilibrium condition is then given by Eq. (\ref{eq:beta}), shown below~\cite{Stone2007} 
\begin{eqnarray}
\label{eq:beta}
    \delta \{ &&\epsilon_B(n,p,\Lambda, \Xi^0,\Xi^-) + \epsilon_e(e)+\epsilon_\mu(\mu)  \nonumber \\
    &&+\Lambda_1\sum n_iq_i+\Lambda_2(\sum_f n_f-n_B) \}=0 \, .
\end{eqnarray}
\begin{figure*}
    \centering
    \includegraphics[scale=1]{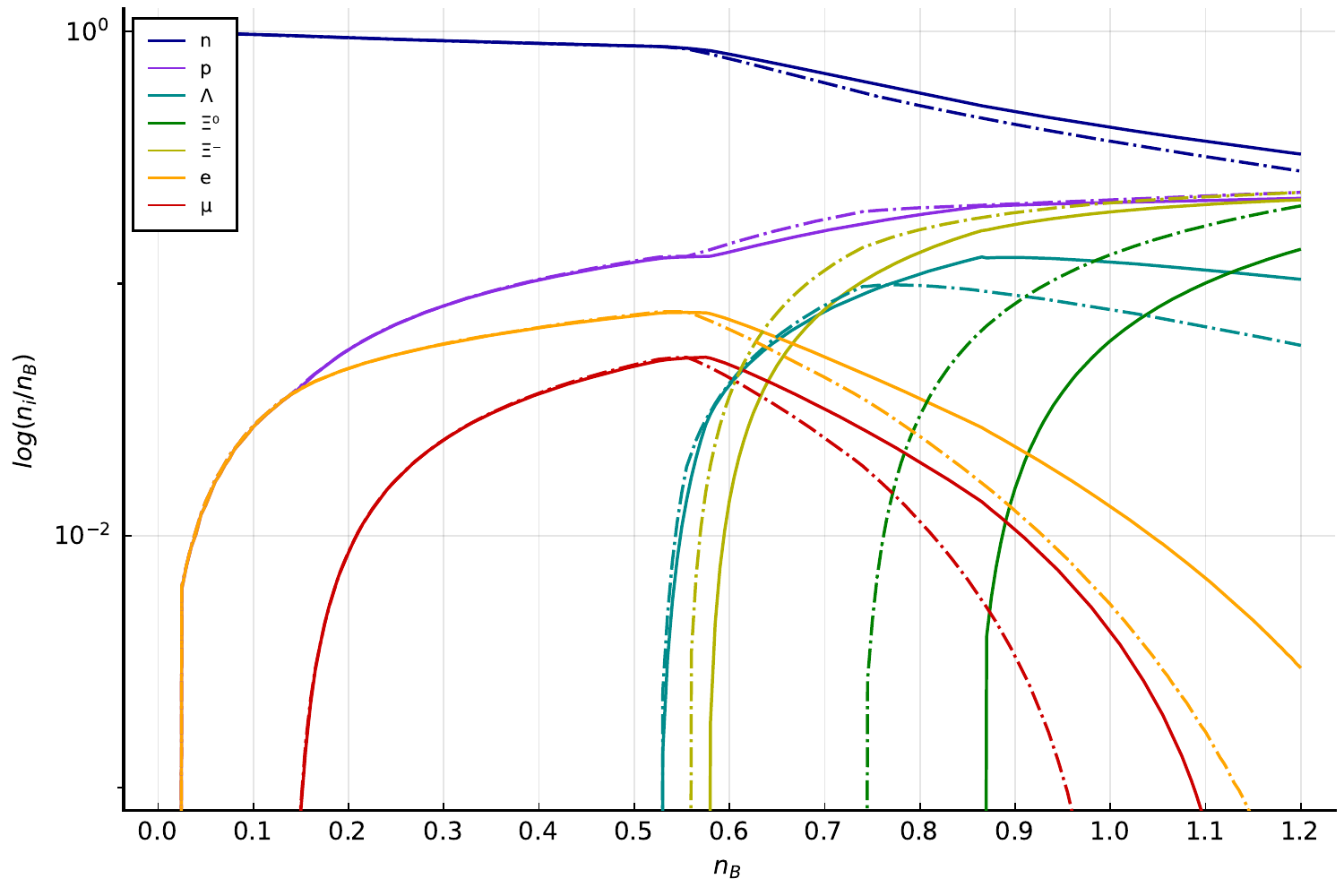}
    \caption{$\beta$-equilibrium was calculated for $0.0<n_B<1.2$ fm$^{-3}$. Species fraction of the nucleons, leptons, and hyperons are shown relative to the total baryon number, $n_B$, for F-QMC (including the effect of overlap). Solid lines indicate $\lambda_3=0.02$ fm$^{-1}$, while the dashed lines are for the case $\lambda_3=0.00$ fm$^{-1}$. Not shown are the curves for F-QMC with no overlap, as they have identical species fractions.}
    \label{fig:specfrac}
\end{figure*}

\section{Results}
\label{sec:Results}
The EoS generated by the QMC model supplemented by the phenomenological overlap term (Eq.~\ref{eq:overlap} ) will be studied either with or without the term cubic in the $\sigma$ field. The latter is labelled $\sigma^3$, with its strength determined by the coefficient $\lambda_3$, which is chosen to be $\lambda_3=0.02$ fm$^{-1}$ or $\lambda_3=0.00$ fm$^{-1}$. The choice $\lambda_3=0.02$ fm$^{-1}$ is motivated by the study of the energies of giant monopole resonances by Martinez~\textit{ et al.}~\cite{Martinez:2018xep}, where it was found that this was the smallest value of $\lambda_3$ capable of producing a value for the nuclear incompressibility compatible with that data.   Where the overlap term is explicitly included, the EoS will be denoted by \textit{Overlap}. When it is not included the EoS will be denoted \textit{No Overlap}. If the derived EoS includes hyperons, then it will be denoted \textit{F-QMC}, while the case where only nucleons are included will be denoted \textit{N-QMC} as a comparison.

Given that the inclusion of $\lambda_3 = 0.02$ fm$^{-1}$ leads to a maximum NS mass that is unacceptably low when hyperons are included, various overlap parameters ($E_0$ and $b$) were explored:
\begin{enumerate}
\label{list:over}
  \item Overlap energy: $E_0=3500$, $4500$, $5500$ MeV
  \item Range parameter: $b=0.4$, $0.5$ fm
\end{enumerate}
The upper limits on both the range and strength parameters were set by the requirement that there be no significant change in the properties of symmetric nuclear matter at saturation density.
Where the parameters are left unspecified in what follows, they were chosen to be $E_0=5500$ MeV and $b=0.5$ fm. These are the preferred choice, in that they lead to an acceptable maximum mass while not altering the nuclear matter parameters.

The bulk properties of the NS are used to test the viability of the model at high density. The properties of interest are the mass-radius relationship and the tidal deformability. For brevity, only PSR J0740+6620 is shown when constraining the QMC EoS. The mass is taken to be $M=2.072^{+0.067}_{-0.066}$ M$_\odot$, with a 68\% interval around the median~\cite{Riley:2021pdl, Fonseca:2021wxt}. GW170817 serves as a constraint for the tidal deformability, $\Lambda_{\textrm{M}}$~\cite{LIGOScientific:2018cki}.

\subsection{The Choice of Parameters}
\label{sec:parameterfixing}
The mass of the $\sigma$ meson is set at 700 MeV, while the masses of the other mesons are taken from their experimental values ($m_\delta=983$ MeV, $m_\rho=770$ MeV, $m_\omega=783$ MeV). In terms of these masses the meson-nucleon coupling strengths in free space are often written as
\begin{eqnarray}
   G_\sigma=\frac{g_\sigma^2}{m_\sigma^2}, \quad
   G_\omega=\frac{g_\omega^2}{m_\omega^2}, \quad
   G_\rho=\frac{g_\rho^2}{m_\rho^2}.
\end{eqnarray}

$G_\delta=3$ fm$^2$ is chosen as the preferred value for the coupling of the $\delta$ field~\cite{Motta:2019tjc,Haidenbauer:1992tn} but for comparison, in Appendix A all calculations are repeated for $G_\delta=0$ fm$^2$. 

In all cases (with and without $\lambda_3$ and $G_\delta$) we require that the bag overlap terms do not alter the properties of nuclear matter at saturation density. Those parameters are fixed at the typical values $n_0=0.16$ fm$^{-3}$, the binding energy per nucleon at $n_0$ is taken to be $E/A = -15.8$ MeV, while the symmetry energy is $S=30$ MeV \cite{Shapiro:1983du, Glendenning:1997wn, Li:2013ola}. As the incompressibility and the slope of the symmetry energy  are typically taken to lie in the range $K_\infty=250\pm50$ MeV \cite{Stone:2014wza, Dutra:2012mb} and $L_0=60\pm20$ MeV \cite{Li:2013ola}. We choose to use $K_\infty=260$ MeV and $L_0=62$ MeV respectively, because, while the relation between the incompressibility and the energies of the giant monopole resonances (GMR) is somewhat complicated~\cite{Sharma:2008uy,Piekarewicz:2002jd}, calculations of the GMR using the QMC EDF tend to favor values of $K_\infty$ at the lower end of this range~\cite{Stone:2014wza}.

As reported by Guichon~\textit{et al.}~\cite{Guichon:2018uew}, the inclusion of $\lambda_3 =0.02$ fm$^{-1}$ lowers $K_\infty$ by about 10\%, leading to the value 260 MeV noted earlier. This remains unchanged for $b = 0.4$ fm for $E_0$ ranging from 3500 to 5500 MeV, while it increases slightly (from 262 to 264 MeV) over this range of $E_0$ for $b=0.5$ fm. So long as $b<0.6$ fm then $K_\infty<300$ MeV is within the acceptable limits. On the other hand, for $\lambda_3 = 0.00$ fm$^{-1}$ the incompressibility is 295 MeV (rising to 298 MeV for $b=0.5$ fm). Physically the range parameter sets the scale at which the extra repulsive force acts in medium. Since saturation density is relatively low, the overlap term has essentially no influence there and thus does not affect the properties of finite nuclei. In NS, the gravitational force compacts the baryonic matter well past saturation, allowing them to eventually overlap~\cite{Ozel:2016oaf}. The extra repulsion induced by baryon overlap stiffens the EoS at supra-nuclear densities. This is explored in 
subsection~\ref{sec:EoS}. 

%%%%%%%%%%%%%%%%%%%%%%%%%%%%%
%%%%%%%%%%%  Species Fraction %%%%%%%%%%%%
\subsection{\texorpdfstring{NS composition under $\beta$-Equilibrium}{NS composition}}
\label{sec:composition}
QMC has previously demonstrated the appearance of only the $\Lambda$ and $\Xi^{0,-}$ hyperons~\cite{Stone:2019blq,Whittenbury:2013wma}. Because of the enhancement of the color hyperfine interaction in-medium~\cite{Guichon:2008zz} and the repulsive three-body force generated by the scalar polarisability, the $\Sigma^{\pm,0}$ baryons experience significant repulsion and are not energetically allowed at densities $n_b\leq1.2$ fm$^{-3}$. The same physical mechanism explains why $\Sigma$-hypernuclei do not exist and also leads to the exclusion of $\Delta$ baryons in NS in this model~\cite{Motta:2019ywl}. 

The species fractions inside a NS, as predicted by F-QMC, are shown in Fig.~\ref{fig:specfrac}. F-QMC predicts no hyperons below 3 $n_0$. The overlap term has no bearing on the species fraction because the repulsion introduced by the overlap is independent of quark content and hence does not discriminate between baryon species (see Eq.~(\ref{eq:overlap})). There is a difference in the appearance of hyperons with (solid) and without (dashed) $\lambda_3$. The $\Xi^{0,-}$ appears slightly later when the term in $\sigma^3$ is present. The relative abundances are also modified.  %(Fig.~\ref{fig:specfrac}). 
\begin{table}
\caption{\label{tb:chemicalpotentials} The chemical potentials, $\mu_i$ in MeV, for each baryon at saturation density, with and without overlap.}
\begin{ruledtabular}
    \centering
    \begin{tabular}{c|ccccc}
     & \multicolumn{5}{c}{$\lambda_3=0.02$ fm$^{-1}$}\\
    F-QMC & n & p & $\Lambda$ & $\Xi^0$ & $\Xi^-$ \\ \hline
    Overlap & 970  & 857  & 1076  & 1300  & 1326 \\
    No Overlap & 970 & 857 & 1076 & 1300 & 1326 \\ \hline
    & \multicolumn{5}{c}{$\lambda_3=0.00$ fm$^{-1}$}\\
    F-QMC & n & p & $\Lambda$ & $\Xi^0$ & $\Xi^-$ \\
    \hline
    Overlap & 970  & 856  & 1080  & 1298  &  1333 \\
    No Overlap &  970 & 856  & 1080  & 1298  & 1333  \\
    
    \end{tabular}
    \end{ruledtabular}
\end{table}

The chemical potentials of the nucleons, hyperons and leptons are unaffected by changes in the overlap parameters. Table~\ref{tb:chemicalpotentials} lists the chemical potentials of the baryon species with and without overlap ($E_0=5500$ MeV and $b=0.5$ fm). We see that the $\Lambda$ experiences an attractive potential of $35-40$ MeV, while the attraction felt by the $\Xi^0$ is considerably smaller. These values are consistent with the fact that the $\Lambda$ is bound in the 1s-state in Pb by around 26 MeV~\cite{Hashimoto:2006aw}, as well as with the recent observation of a $\Xi$ weakly bound to  $^{14}$N~\cite{Nakazawa:2015joa,Shyam:2019laf,Guichon:2008zz}.

%%%%%%%%%%%%%%%%%%%%%%%%%%%%%%%
%%%%%%%%%%%%%%%%%%%%%%%%%%%%%%%%%%%
\subsection{QMC EoS}
\label{sec:EoS}
\begin{figure*}
    \centering
    \includegraphics[scale=1]{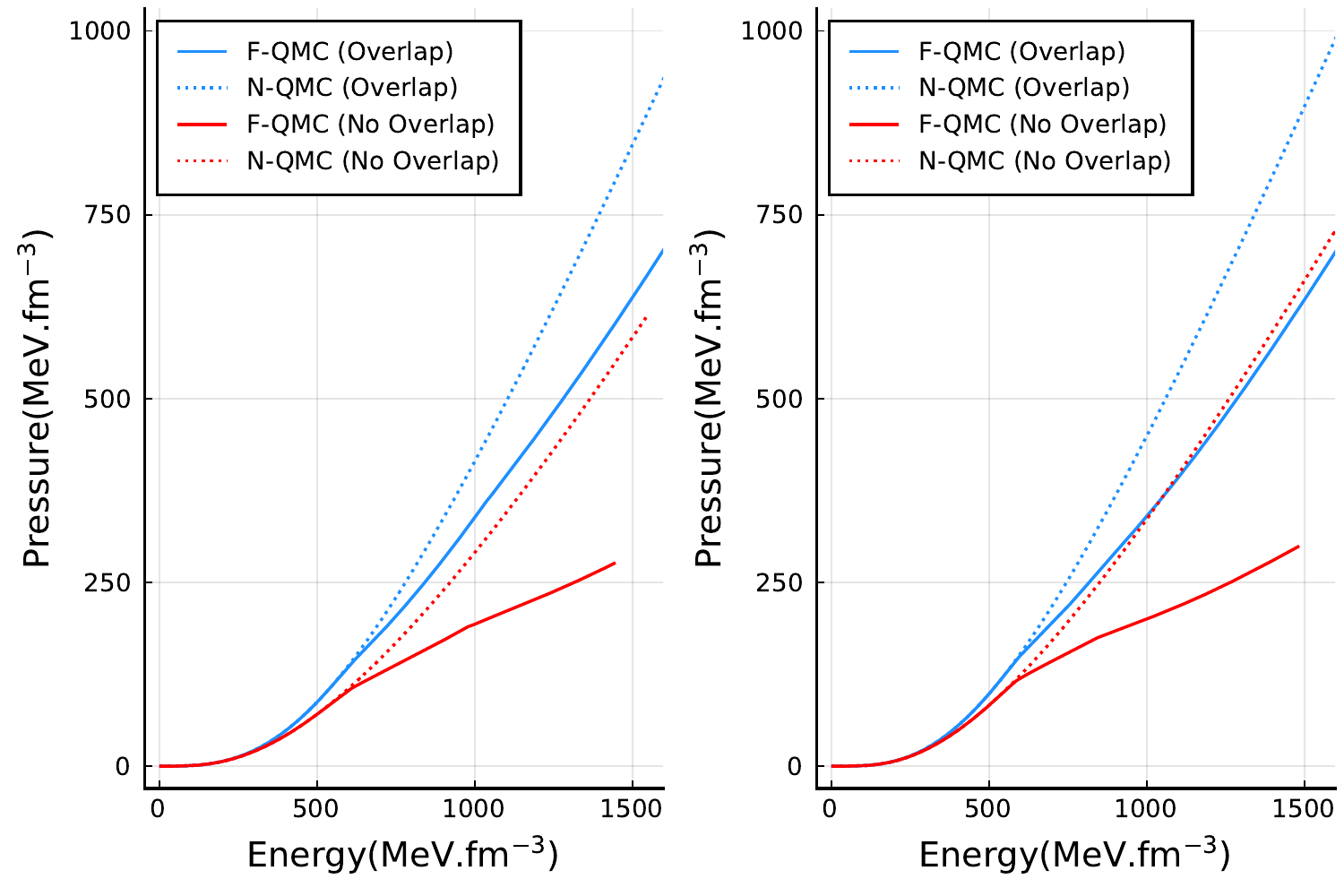}
    \caption{The EoS for the models considered here, where N-QMC includes only nucleons and F-QMC includes hyperons. The left panel corresponds to $\lambda_3=0.02$ fm$^{-1}$, whilst the right panel is $\lambda_3=0.00$ fm$^{-1}$.}
    \label{fig:EoS}
\end{figure*}
\begin{figure*}
    \centering
    \includegraphics[scale=1]{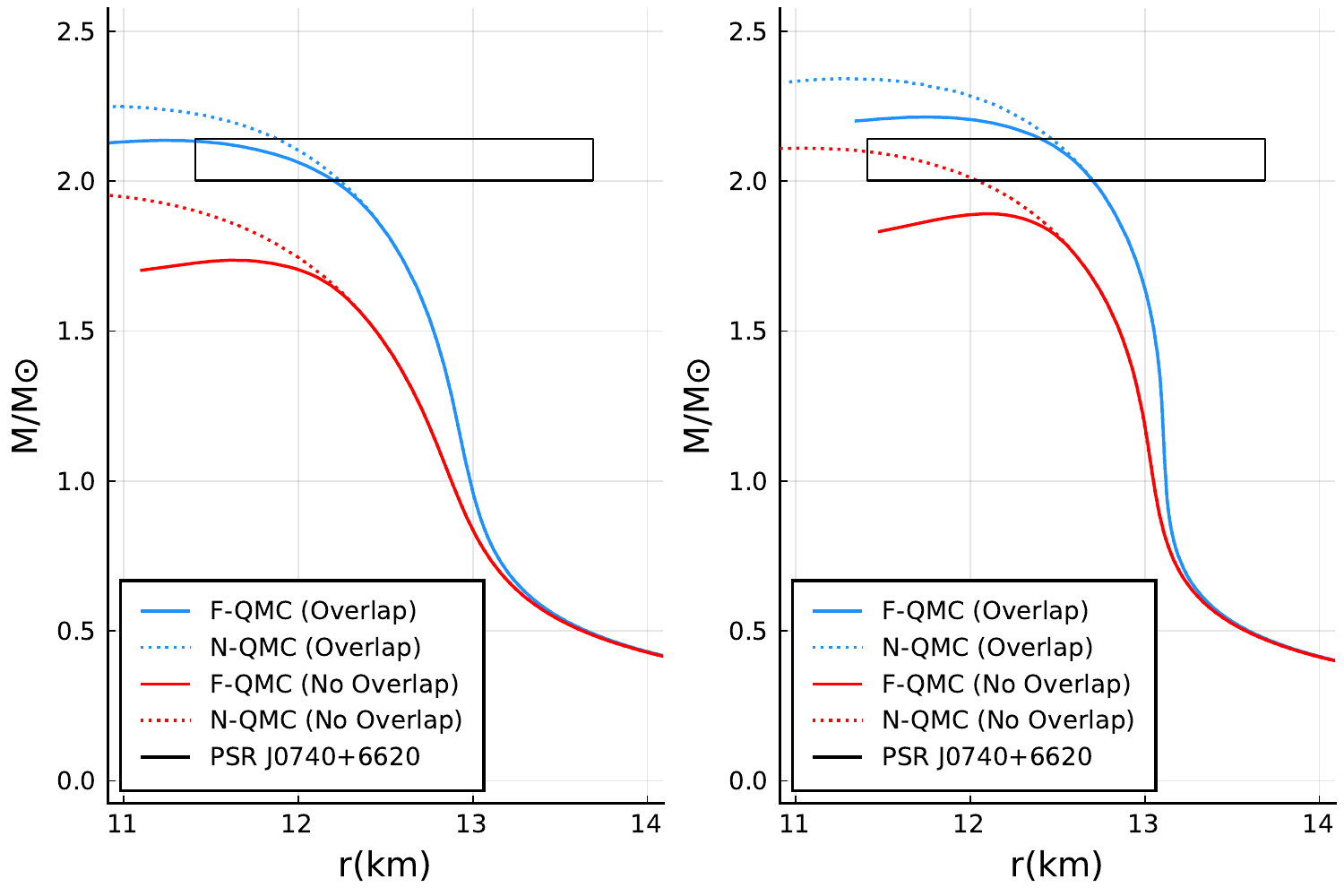}
    \caption{The mass-radius sequence of stars computed from QMC EoS shown in Fig. \ref{fig:EoS}. The box constraint (black) is PSR J0740+6620, extracted from Ref.~\cite{Riley:2021pdl}. The left panel is $\lambda_3=0.02$ fm$^{-1}$, whilst the right panel is $\lambda_3=0.00$ fm$^{-1}$.} 
    \label{fig:MR}
\end{figure*}
The low density crustal region enveloping the core of a NS is populated by nuclei with increasing neutron excess~\cite{Chamel:2015oqa,Yakovlev:2000jp,Antic:2020zuk}. Here the QMC EoS, which is an appropriate description of nuclear matter, is matched onto the low density EoS provided by Hempel and Schaffner-Bielich  \cite{Hempel:2009mc, Hempel:2011mk}.  The Hempel and Schaffner-Bielich model is a relativistic mean field model for interacting nucleons which takes into account excluded volume effects. In what follows the QMC EoS is matched to the crust region at $n\approx0.7$ $n_0$, which is appropriate in describing the transition of nuclei to nuclear matter at the crust-core boundary.

The derived F-QMC (solid) and N-QMC (dashed) EoS, with crust, are shown in Fig.~\ref{fig:EoS}. Note that in all figures, unless otherwise stated, the overlap case corresponds to $b=0.5$ fm and $E_0 = 5500$ MeV. As expected, the hyperons soften the EoS when compared to that for nucleons only. While the inclusion of the overlap term has no influence on the nuclear matter parameters at saturation density, it is clear that the EoS is significantly stiffer at high density. In Fig.~\ref{fig:EoS} we see that the effect of the overlap term becomes considerable at energy densities of order $250-350$ MeV-fm$^{-3}$, or $n>2$ $n_0$. Furthermore, the degree of softening induced by the hyperons is reduced at higher densities. Comparing the two panels in Fig.~\ref{fig:EoS}, the cubic term in  $V(\sigma)$ acts to soften the EoS, whether hyperons are included or not.

In order to make the QMC EoS generated here (without crust) widely available, an analytic function has been fitted to the F-QMC EoS with $\lambda_3=0.02$ fm$^{-1}$ and $\lambda_3=0.00$ fm$^{-1}$. This is valid for the energy density between $0-1600$ MeV-fm$^{-3}$ but must be matched to a crust EoS for $n<0.70$ $n_0$, corresponding to an energy density $\epsilon\approx105$ MeV-fm$^{-3}$. Eq.~(\ref{eq:powerlaw}) takes the argument for energy density in MeV-fm$^{-3}$ and gives the pressure in MeV-fm$^{-3}$.
\begin{eqnarray}
    \label{eq:powerlaw}
    P(\epsilon)=N_1\epsilon^{p_1}+N_2\epsilon^{-p_2}.
\end{eqnarray}

The error computed is given by Eq.~(\ref{eq:error}) and the parameters are summarised in Table~\ref{tb:AnalyticEoS} below, with the energy density split into different regions. Eq.~(\ref{eq:powerlaw}) is not suitable for computing the speed of sound (Eq.~(\ref{eq:sound})) as the domain boundaries do not precisely correspond to the appearance of new species. 
\begin{eqnarray}
    \label{eq:error}
    \textrm{RMSE (\%)} = \sqrt{\frac{1}{N} \sum_i^N \frac{(x_i-y_i)^2}{y_i^2} }\times100.
\end{eqnarray}
\begin{table}
\caption{\label{tb:AnalyticEoS} Parameters for Eq.~(\ref{eq:powerlaw}), corresponding to the F-QMC EoS with $\lambda_3=0.02$ fm$^{-1}$ and $\lambda_3=0.00$ fm$^{-1}$. The domains for the energy density, $\epsilon$ (MeV-fm$^{-3}$), have been split as denoted by the left hand column.}
\begin{ruledtabular}
    \centering
    \begin{tabular}{c|ccccc}
    & \multicolumn{5}{c}{F-QMC with $\lambda_3=0.02$ fm$^{-1}$} \\
    $\epsilon$ & $N_1$ & $p_1$ & $N_2$ & $p_2$ & RMSE \\ \hline
    %domain 0
    0-34 &  7.733$\times10^{-4}$ & 1.203 & - & - & 1.49\% \\ 
    %domain 1
    35-90& 6.171$\times10^{-7}$ & 3.043 & 1.578 & 1.128 & 0.64\% \\ 
    %domain 2
    91-133&  3.309$\times10^{-7}$& 3.186  & - & - & 0.10\% \\ 
    %domain 3
    134-298&  1.260$\times10^{-6}$ &2.921 & - & - & 1.09\% \\ 
    %domain 4
    299-550&  5.884$\times10^{-6}$ &2.656 & - & - & 1.16\% \\ 
    %domain 5
    551-620&  8.387$\times10^{-5}$& 2.234 & - & - & 0.13\% \\ 
    %domain 6
    621-1021&  1.730$\times10^{-3}$ & 1.764  & - & - & 0.10\% \\
    %domain 7
    1022-1600& 8.269$\times10^{-3}$ &1.539 & - & - & 0.11\% \\
      \hline
    & \multicolumn{5}{c}{F-QMC with $\lambda_3=0.00$ fm$^{-1}$} \\
    $\epsilon$  & $N_1$ & $p_1$ & $N_2$ & $p_2$ & RMSE \\ \hline
    %domain 0
    0-24  & 9.725$\times10^{-4}$ & 1.183 & - & - & 4.04\% \\
    %domain 1
    25-90 & 6.234$\times10^{-8}$ & 3.498 & 1.549$\times10^{-2}$ & -0.3064 & 1.10\% \\ 
    %domain 2
    91-162 & 1.376$\times10^{-7}$ & 3.355 & - & - & 0.66\% \\ 
    %domain 3
    163-299& 6.243$\times10^{-7}$ & 3.064& - & - & 1.02\% \\ 
    %domain 4
    300-549& 7.150$\times10^{-6}$ & 2.645 & - & - & 1.59\% \\ 
    %domain 5
    550-595& 1.143$\times10^{-4}$ & 2.204 & - & - & 0.06\% \\ 
    %domain 6
    596-861& 4.517$\times10^{-3}$ & 1.629 & - & - & 0.05\% \\ 
    %domain 7
    862-1600& 8.295$\times10^{-3}$ & 1.538 & - & - & 0.25\% \\

    \end{tabular}
    \end{ruledtabular}
\end{table}
%

%%%%%%%%%%%%%%%%%%%%%%%%%%%%%%%%%%%%%%%%
%%%%%%%% MASS and Radius %%%%%%%%%%%
\subsection{NS Bulk Properties}
The Tolman-Oppenheimer-Volkoff (TOV) equation was used to compute the mass and radius of the NS. Assuming that the NS is non-rotating and spherically symmetric, the TOV equation is 
\begin{eqnarray}
\label{eq:TOV}
    \frac{dp}{dr}=-\frac{\left[p(r)+\epsilon(r)\right]\left[M(r)+4\pi r^3p(r)\right]}{r\left(r-2M(r)\right)} \, ,
\end{eqnarray}
where
\begin{eqnarray}
\label{eq:massstar}
    M(r)&=4\pi\int_0^r \epsilon(r')(r')^2dr' \, .
\end{eqnarray}
The central pressure is chosen at $r=0$ and integrated outwards until $p(R)=0$, where $R$ is the final radius of the star. This process is repeated for different central pressures to form the sequence of stars plotted in Fig.~\ref{fig:MR}. The box denotes the constraint corresponding to pulsar PSR J0740+6620, $M=2.072^{+0.067}_{-0.066}$ M$_\odot$ and $R=12.39^{+1.30}_{-0.98}$ km. 
The total baryon number is given by
\begin{eqnarray}
    A=\int_0^R \frac{4\pi r^2n_B(r)}{(1-\frac{2 G m(r)}{r})^\frac{1}{2}}dr
    \, .
\end{eqnarray}

GW 170817 is a binary system with a total mass of $2.73^{+0.04}_{-0.01}$ M$_\odot$~\cite{LIGOScientific:2018cki}. The component masses in the low spin case are $m_1\in(1.36, 160)$ M$_\odot$ and $m_2\in(1.16, 1.36)$ M$_\odot$. The tidal deformability may be computed as
\begin{eqnarray}
    \label{eq:tidal}
    \Lambda_M=\frac{2}{3}k_2 \left(\frac{R}{M} \right)^5 \, .
\end{eqnarray}
The dimensionless constant, $k_2$, is the tidal love number and the full equation is given in Ref.~\cite{Meng:2021ijp,Chatziioannou:2020pqz}. Eq.~(\ref{eq:tidal}) gives the one sided tidal deformation, which is constrained by GW170817 for a $1.4$ M$_\odot$ star to lie in the range $\Lambda_{1.4}=190^{+390}_{-120}$~\cite{LIGOScientific:2018cki}. This is shown as a black line in Fig.~\ref{fig:Tidal}. However, we note that other work has reported that the upper limit on $\Lambda_{1.4}$ could be as large as $800$ MeV~\cite{Kim:2018aoi}.

%%%%%%%%%%%%%%%%%%%%%%%%%%%%%%%%%%%%%%%%%%%%%%%%%%%%
\subsubsection{Mass-Radius Relation}
Table~\ref{tb:MRtable} summarises the properties of a NS with maximum mass, $M_{max}$, predicted by QMC. All entries are for the case where hyperons are included, unless otherwise indicated. The overlap parameters alter the NS properties in predictable ways. Increasing $E_0$ has mild effects on the maximum mass of the star, with little change to its radius (see Fig. \ref{fig:MR}). The range parameter $b$, however, raises the mass substantially as it is increased. For $b=0.4$ fm and $\lambda_3 = 0.02$ fm$^{-1}$, the maximum mass is predicted to be $M_{max}<2$ M$_\odot$, which is unsatisfactory. 
\begin{table}
\caption{\label{tb:MRtable} Macroscopic properties of the NS (including hyperons unless otherwise indicated) are computed with variations of the overlap parameters. The range parameter, $b$, and overlap energy, $E_0$, have units of fm and MeV, respectively. The results summarise the maximum mass ($M_{max}$, $M_\odot$), total baryon number ($A$, $10^{57}$) and central number density ($n_c$, fm$^{-3}$), central pressure ($P_c$, MeV-fm$^{-3}$), and central energy density ($\epsilon_c$, MeV-fm$^{-3}$) for each parameter set used.}
\begin{ruledtabular}
    \centering
    \begin{tabular}{cc|ccccc}
    & & \multicolumn{5}{c}{$\lambda_3=0.02$ fm$^{-1}$}\\
    $b$&$E_0$&$M_{max}$&A&$n_c$& $P_c$ &$\epsilon_c$ \\ \hline
    0.4 & 3500 & 1.77 & 2.41 & 1.01 & 256 & 1191 \\
    0.4 & 4500 & 1.78 & 2.43 & 1.02 & 268 & 1205 \\
    0.4 & 5500 & 1.79 & 2.45 & 1.03 & 280 & 1219 \\ \hline
    0.5 & 3500 & 2.02 & 2.82 & 1.02 & 408 & 1260 \\
    0.5 & 4500 & 2.08 & 2.92 & 1.01 & 448 & 1258 \\
    0.5 & 5500 & 2.14 & 3.02 & 1.00 & 492 & 1267 \\ \hline
    0.5\footnote{\label{ft:Nucleonsoverlap}N-QMC (Overlap)} & 5500 & 2.25 & 3.22 & 1.00 & 680 & 1314 \\
    -\footnote{\label{ft:hyperonsNoOverlap}F-QMC (No Overlap)} & 0 & 1.74 & 2.36 & 0.961 & 213 & 1111 \\
    -\footnote{\label{ft:NucleonsNoOverlap}N-QMC (No Overlap)} & 0 & 1.96 & 2.74 & 1.15 & 559 & 1461 \\ \hline
    & & \multicolumn{5}{c}{$\lambda_3=0.00$ fm$^{-1}$}\\
    $b$&$E_0$&$M_{max}$&A&$n_c$& $P_c$ &$\epsilon_c$ \\ \hline
    0.4 & 3500 & 1.91 & 2.63 & 0.883 & 221 & 1030 \\
    0.4 & 4500 & 1.92 & 2.64 & 0.900 & 231 & 1055 \\
    0.4 & 5500 & 1.92 & 2.65 & 0.915 & 242 & 1078 \\ \hline
    0.5 & 3500 & 2.11 & 2.97 & 0.906 & 336 & 1102 \\
    0.5 & 4500 & 2.17 & 3.05 & 0.923 & 387 & 1144 \\
    0.5 & 5500 & 2.21 & 3.14 & 0.903 & 406 & 1123 \\ \hline
    0.5 $^{\textrm{a}}$ & 5500 & 2.34 & 3.37 & 0.934 & 643 & 1222 \\
    - $^{\textrm{b}}$ & 0 & 1.89 & 2.60 & 0.867 & 201 & 1005 \\
    - $^{\textrm{c}}$ & 0 & 2.11 & 2.97 & 1.03 & 529 & 1305 \\
    
    \end{tabular}
    \end{ruledtabular}
\end{table}

Table~\ref{tb:Central} reflects the central properties of different mass NS predicted for F-QMC with overlap. The central density, pressure and energy densities are all greater when $\lambda_3=0.02$ fm$^{-3}$, for all masses. For a star of mass $M=1.4$ M$_\odot$, the number density is lower than the threshold density for hyperons and hence there are no hyperons in these stars. However, as the density increases the star's core is then populated by hyperonic matter. 
\begin{table}
\caption{\label{tb:Central} The central number density ($n_c$, fm$^{-3}$), pressure ($P_c$, MeV-fm$^{-3}$) and energy density ($\epsilon_c$, MeV-fm$^{-3}$) for different mass stars (M$_\odot$) predicted by F-QMC with overlap ($E_0=5500$ MeV, $b=0.5$ fm).}
\begin{ruledtabular}
    \centering
    \begin{tabular}{cccc}
    \multicolumn{4}{c}{$\lambda_3=0.02$ fm$^{-1}$}\\
    Mass & $n_c$ & $P_c$ & $\epsilon_c$\\ \hline
    1.0 & 0.341 & 30 & 336 \\
    1.4 & 0.427 & 59 & 432 \\
    1.6 & 0.482 & 86 & 496 \\
    1.8 & 0.547 & 124 & 576 \\
    2.0 & 0.663 & 196 & 733 \\ \hline
    \multicolumn{4}{c}{$\lambda_3=0.00$ fm$^{-1}$}\\
    Mass & $n_c$ & $P_c$ & $\epsilon_c$\\ \hline
    1.0 & 0.317 & 27 & 311 \\
    1.4 & 0.395 & 53 & 397 \\
    1.6 & 0.438 & 74 & 447 \\
    1.8 & 0.487 & 102 & 507 \\
    2.0 & 0.564 & 154 & 607 \\
    \end{tabular}
    \end{ruledtabular}
\end{table}

In Fig.~\ref{fig:MR}, the mass-radius curve shows that the overlap term is essential in predicting a heavy NS, $M>2$ M$_\odot$, once the incompressibility is reduced to the preferred range (i.e., with $\lambda_3=0.02$ fm$^{-1}$). Without the overlap term, the mass of the star is significantly lower. The inclusion of the $\sigma^3$ term acts to reduce the radius and lower the star's mass. The mass reduction is caused by the additional scalar meson attraction, with a consequent softening of the EoS. For $\lambda_3=0.02$ fm$^{-1}$, the radius of the star slightly increases as the mass decreases from $1.5$ M$_\odot$ to $1.0$ M$_\odot$, in contrast to the case $\lambda_3=0.0$ fm$^{-1}$, where there are no significant changes to the radius. In the phenomenologically interesting region, $M \, \approx \, 1.4 \, M_\odot$, the radius of the star is significantly lower when $\lambda_3 = 0.02$ fm$^{-1}$.

The presence of hyperons reduces the maximum mass, as well as increasing the radius at maximum mass. From Fig.~\ref{fig:MR} we see that the overlap term decreases the radius at maximum mass for F-QMC, whereas for N-QMC the radius is increased at maximum mass when the overlap term is present. 
%%%%%%%%%%%%%%%%%%%%%%%%%%%%%%%%%%%%%%%%%%%
%%%%%%%      TIDAL    %%%%%%%%%%%%%
%%%%%%%%%%%%%%%%%%%%%%%%%%%%%%%%%%%%%%%%%%%
%\subsubsection{Tidal Deformation}
%\label{sec:Tidal}

\begin{figure*}
    \centering
    \includegraphics[scale=1]{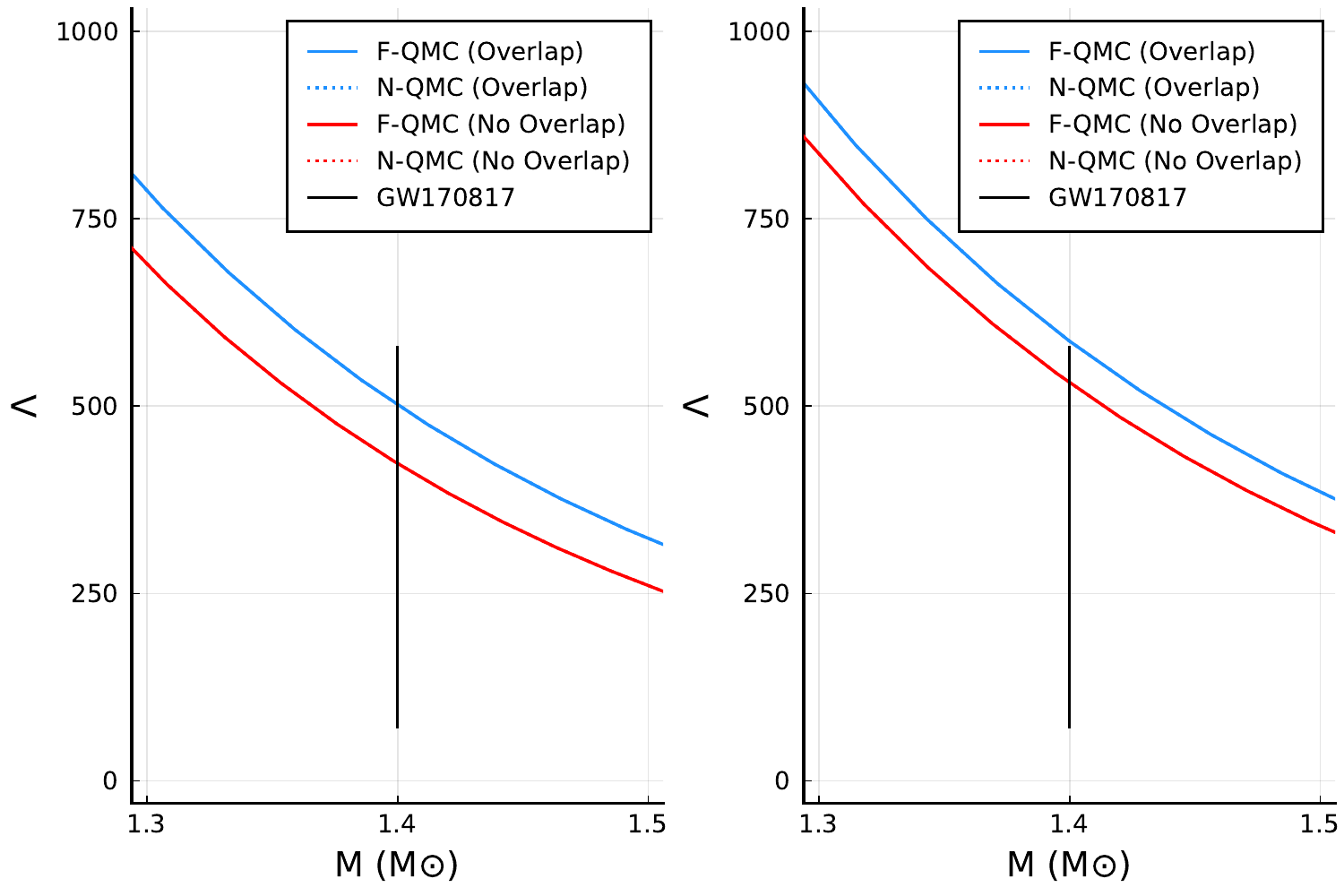}
    \caption{Tidal deformability for the QMC EoS. Given that no hyperons appear in a star of mass 1.4 M$_\odot$ the solid (F-QMC) and dashed (N-QMC) lines overlap. GW170817 is shown by the solid black line and extracted from~\cite{LIGOScientific:2018cki}. The left panel is $\lambda_3=0.02$ fm$^{-1}$, whilst the right panel is $\lambda_3=0.00$ fm$^{-1}$.}
    \label{fig:Tidal}
\end{figure*}

Figure~\ref{fig:Tidal} illustrates the tidal deformability for the QMC EoS. The dashed line for the nucleon only case cannot be distinguished from F-QMC because the QMC model predicts no hyperons in a $1.4$ M$_\odot$ star. 
%%%%%%%%%%%%%%%%%%%
%%%%  sound speed  %%%%%%%%%%
\subsection{Speed of sound}
\label{sec:speed of sound}
\begin{figure*}
    \centering
    \includegraphics[scale=1]{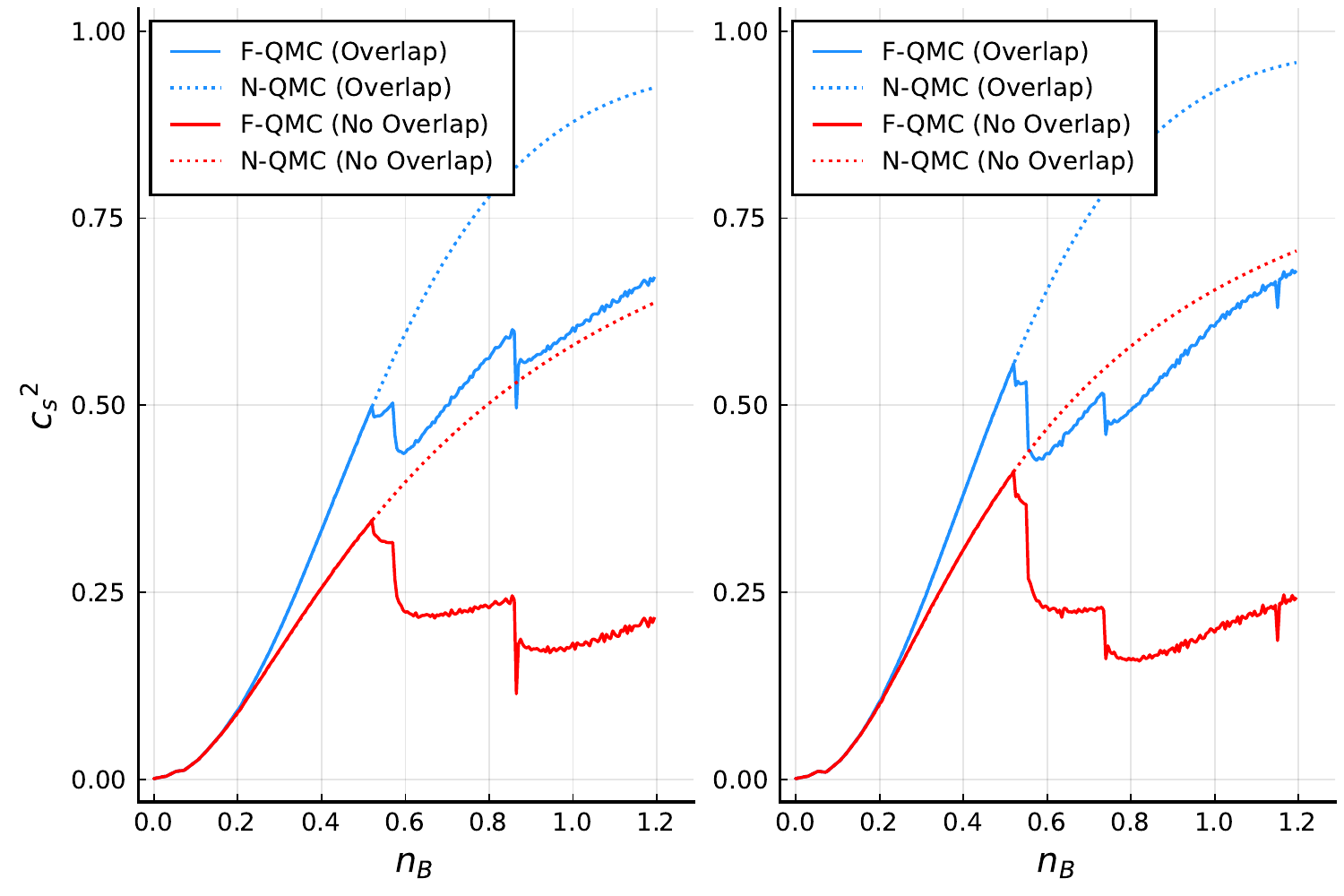}
    \caption{The speed of sound ($c_s$) is given as a fraction of the speed of light for $b=0.5$ fm and $E_0 = 5500$ MeV. The number density $n_B$ is given in fm$^{-3}$. The blue and red lines correspond to the cases with and without overlap, respectively, while the solid and dashed lines show the full octet and nucleons-only cases. The left panel corresponds to  $\lambda_3=0.02$ fm$^{-1}$, whilst the right panel is $\lambda_3=0.00$ fm$^{-1}$.}
    \label{fig:Sound}
\end{figure*}

In the absence of direct observations of the composition of the core of a NS, theoretical calculations of the speed of sound (equivalently, polytropic index, $\gamma$) do offer valuable insights. It has previously been suggested that non-trivial changes in the EoS correspond to phase changes from hadronic matter to either quark matter~\cite{Annala:2019puf} or the threshold of creation of hyperons~\cite{Motta:2020xsg}.

The EoS at low and extremely high densities have been extensively studied in effective field theory ($n<1.1$ $n_0$) and perturbative QCD ($n>40$ $n_0$), respectively. Annala~\textit{et al.}~\cite{Annala:2019puf} suggest that for stars with $M=1.4$ M$_\odot$, hadronic nuclear theories are suitable in predicting the EoS giving rise to canonical mass stars. The QMC model is consistent with their EoS in that region. However, for $M>2$ M$_\odot$, Annala~\textit{et al.} suggest that the central density becomes so large that the cores of the stars may be populated by deconfined quark matter and gluons~\cite{Annala:2019puf}. Quark matter, being conformal and scale invariant, would then have a speed of sound, $c_s^2=\frac{1}{3}$, approaching logarithmically from below as the density increases. On the other hand, it has been shown that this is not a distinct feature of conformal matter~\cite{Motta:2020xsg,Stone:2019blq}. Since the QMC model does not contain any elements of deconfined quarks, $c^2_s<\frac{1}{3}$ approaching from below, may also be interpreted as the creation of hyperons.
\begin{eqnarray}
    c^2_s= \frac{dP}{d\epsilon}.
    \label{eq:sound}
\end{eqnarray}

Figure~\ref{fig:Sound} shows the three sudden changes in $c_s^2$ which occur at those number densities where the different species of hyperons first appear (see Fig.~\ref{fig:specfrac}). In order of appearance, these are the $\Lambda$, $\Xi^-$ and $\Xi^0$. Without the overlap term (red lines), the results reflect those report by earlier for the QMC model~\cite{Motta:2020xsg,Stone:2019blq}. However, when the overlap terms are included (blue lines), $c^2_s$ can be as large as 0.5 or more. This is consistent with pure hadronic matter, as cited in~\cite{Annala:2019puf}. 

\section{Conclusion}
\label{sec:Conclusion}
The excitement of studying neutron stars is that they contain the most dense matter in the Universe. Thus they may be expected to yield insights into the equation of state (EoS) of strongly interacting matter at densities inaccessible in any other way.

Relativistic descriptions of nuclear matter typically lead to higher values of the incompressibility of nuclear matter, $K_\infty$ , than non-relativistic Skyrme forces. While the connection between the incompressibility and the energies of giant monopole resonances (GMR) is complex, there is a tendency for lower values of $K_\infty$ to be preferred. The energy density functional derived within the QMC model requires a small cubic term involving the scalar field in order to reproduce the observed GMR energies. This in turn lowers the maximum masses of the neutron stars generated by the model.

In order to solve that problem, we have explored the impact on the EoS of dense matter and the properties of neutron stars by introducing a phenomenological repulsive contribution to the energy density as the degree to which the baryons overlap increases. This new term is designed such that it does not alter the properties of nuclear matter at saturation density.

The effects of this new ``overlap’’ term are indeed to generate neutron stars with a maximum mass above 2.1 M$_\odot$, when hyperons are included. The radii of stars around 1.4 M$_\odot$ lie just below 13 km, which is the upper end of the range preferred by the analysis of the gravitational wave data from GW170817. The calculated values of the tidal deformability also lie within the bounds determined from that data. The one qualitative change from earlier work is that in contrast with the values of $c_s^2$ found with hyperons, which lay below $\frac{1}{3}$, with the overlap term they become as large as 0.6. Such large values are consistent with those expected in hadronic models, in contrast with the lower range anticipated for quark matter.

\section*{Acknowledgements}
We are pleased to acknowledge Jirina Stone for a careful reading of this manuscript and a number of helpful comments. This work was supported by the University of Adelaide and the Australian Research Council through a grant to the ARC Centre of Excellence for Dark Matter Particle Physics (CE200100008).

\appendix

\section{\texorpdfstring{\\QMC overlap with G$_\delta$=0}{Gdelta=0}}

The results (section \ref{sec:Results}) are replicated with no iso-vector scalar meson, i.e., $G_\delta=0.0$ fm$^2$. As shown by Motta~\textit{et al.}, the isovector $\delta$ increases the maximum mass only marginally but changes the radius significantly~\cite{Motta:2019tjc}. For brevity, not all the results are included; only the preferred values of the overlap parameters are shown ($E_0=5500$ MeV and $b=0.5$ fm). %The overlap parameters are chosen to take the preferred values, $E_0=5500$ MeV and $b=0.5$ fm. 
One finds that the largest effect of the change in $G_\delta$ is that the slope of the symmetry energy, $L$, decreases to 53 MeV (52 MeV for $\lambda_3 = 0.00$ fm$^{-1}$).
%
%
%\begin{table}
%\caption{\label{tb:deltaNP} The properties of nuclear matter for %$G_\delta=0$, with and without overlap. The overlap parameters %are at their preferred values of $E_0=5500$MeV and $b=0.5$fm.}
%\begin{ruledtabular}
%    \centering
%    \begin{tabular}{c|ccccc}
%            & \multicolumn{5}{c}{$\lambda_3=0.02 %\textrm{fm}^{-1}$}\\
%            & $n_0$ & $\mathcal{E}$ & $S$& $K_\infty$  & $L$  \\ 
%            & (fm$^{-3}$) & (MeV) & (MeV) & (MeV) &(MeV) \\ %\hline
%            No Overlap & 0.16 & -15.8 & 30 & 259 & 53 \\
%            Overlap & 0.16 & -15.8 & 30 & 263 & 53 \\
%             \hline
%            & \multicolumn{5}{c}{$\lambda_3=0.00 %\textrm{fm}^{-1}$}\\
%            & $n_0$ & $\mathcal{E}$ & $S$& $K_\infty$  & $L$  \\
%            & (fm$^{-3}$) & (MeV) & (MeV) & (MeV) &(MeV) \\ %\hline
%            No Overlap &  0.16 & -15.8 & 30 & 296 & 52 \\
%            Overlap & 0.16 & -15.8 & 30 & 300 & 52 \\
%    \end{tabular}
%    \end{ruledtabular}
%\end{table}
%The couplings may be found on table \ref{tb:couplings}. 

\begin{figure*}
    \centering
    \includegraphics[scale=1]{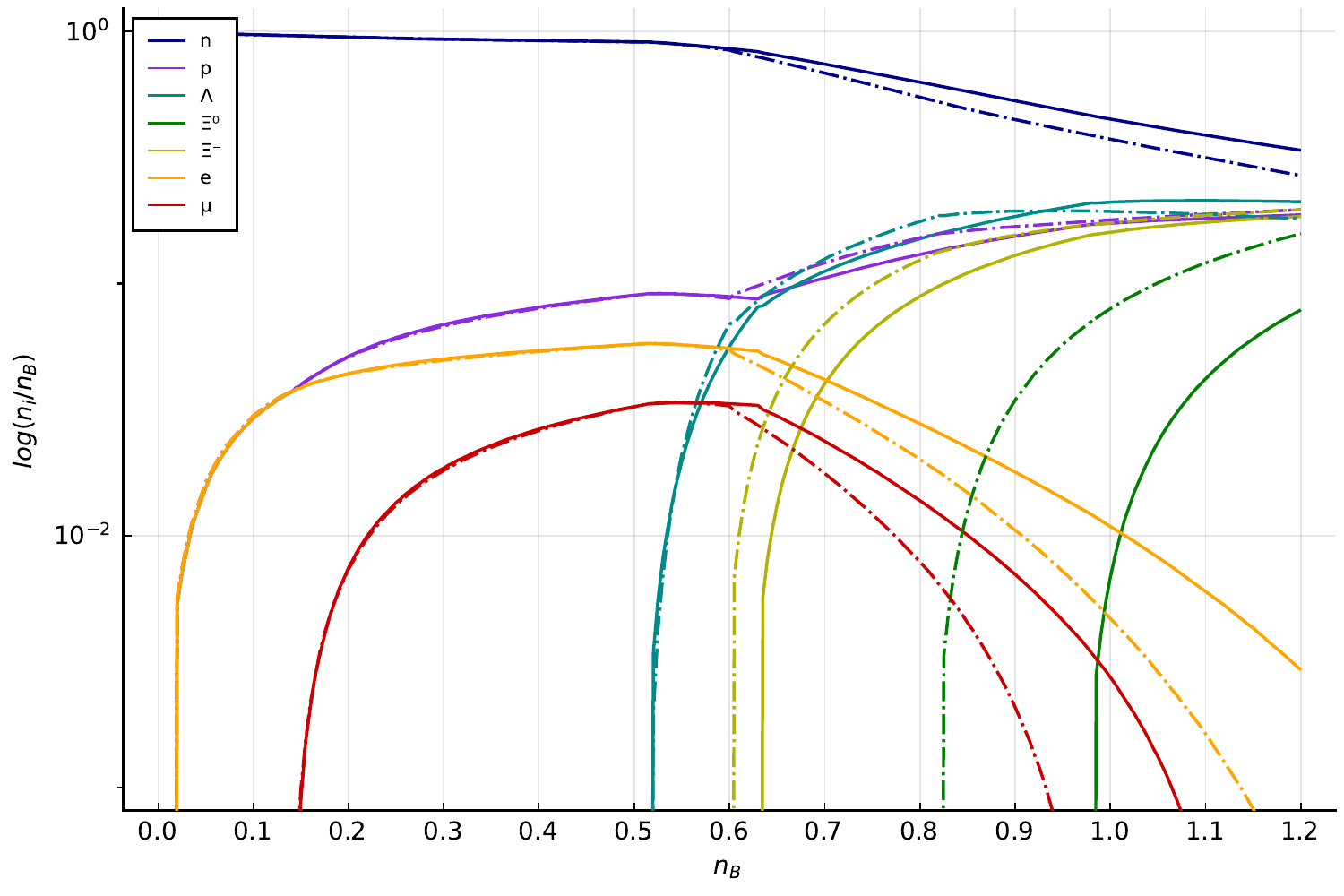}
    \caption{$\beta$-equilibrium was calculated for $0.0<n_B<1.2$ fm$^{-3}$ with $G_\delta=0$ fm$^2$. Only the F-QMC with overlap species fraction is shown with the solid line corresponding the $\lambda_3=0.02$ fm$^{-1}$ and the dashed line showing the case for $\lambda_3=0.0$ fm$^{-1}$. The no overlap case is not shown as it produces identical species fractions.}
    \label{fig:deltaSF}
\end{figure*}

The species fractions are given in Fig.\ref{fig:deltaSF}. The $\Lambda$ appears sooner without the $\delta$ mesons, whilst the $\Xi^{0,-}$ both appear later. 

\begin{figure*}
    \centering
    \includegraphics[scale=1]{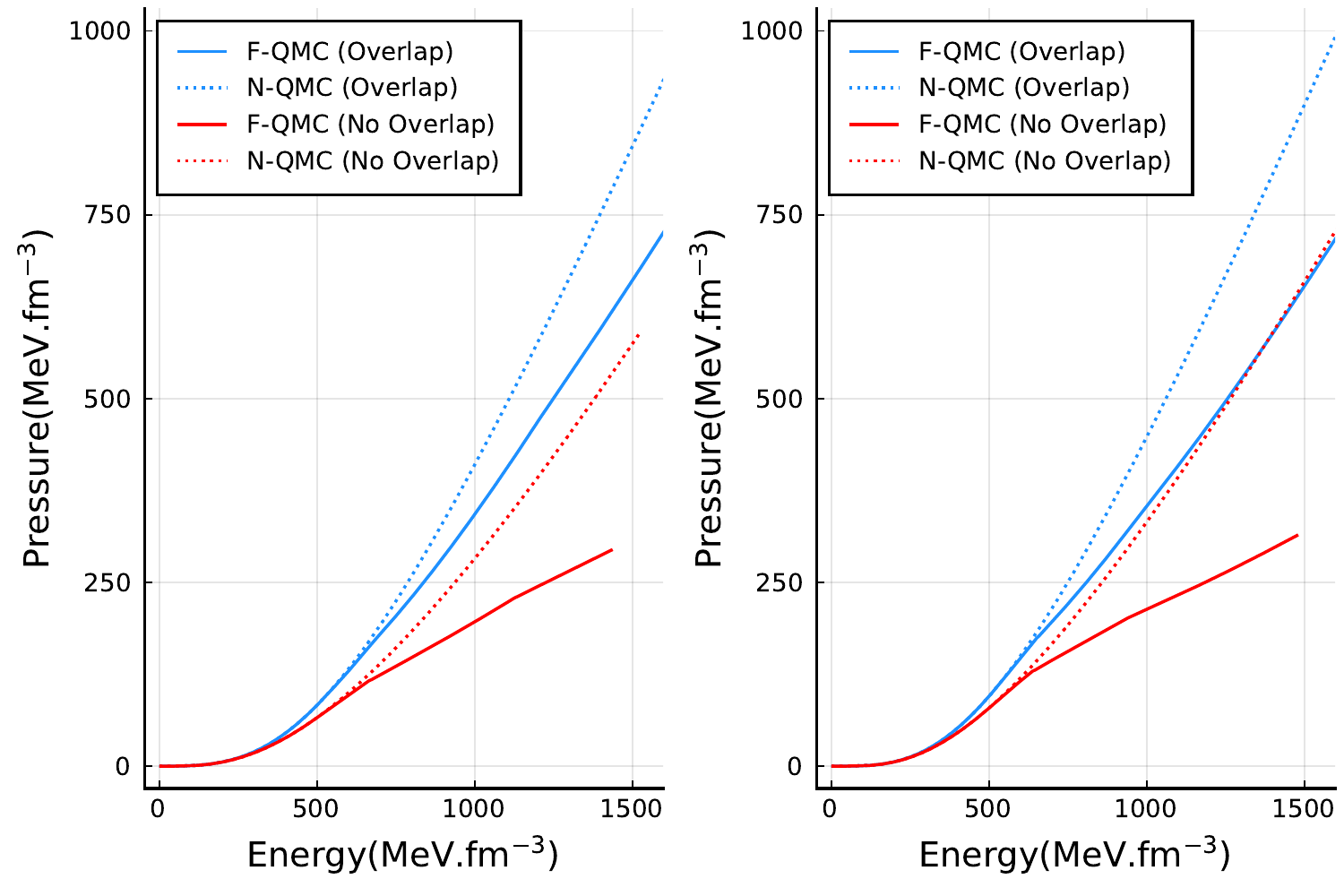}
    \caption{QMC EoS for $G_\delta=0$ fm$^2$. The left panel is $\lambda_3=0.02$ fm$^{-1}$, whilst the right panel is $\lambda_3=0.00$ fm$^{-1}$.}
    \label{fig:deltaEoS}
\end{figure*}

\begin{figure*}
    \centering
    \includegraphics[scale=1]{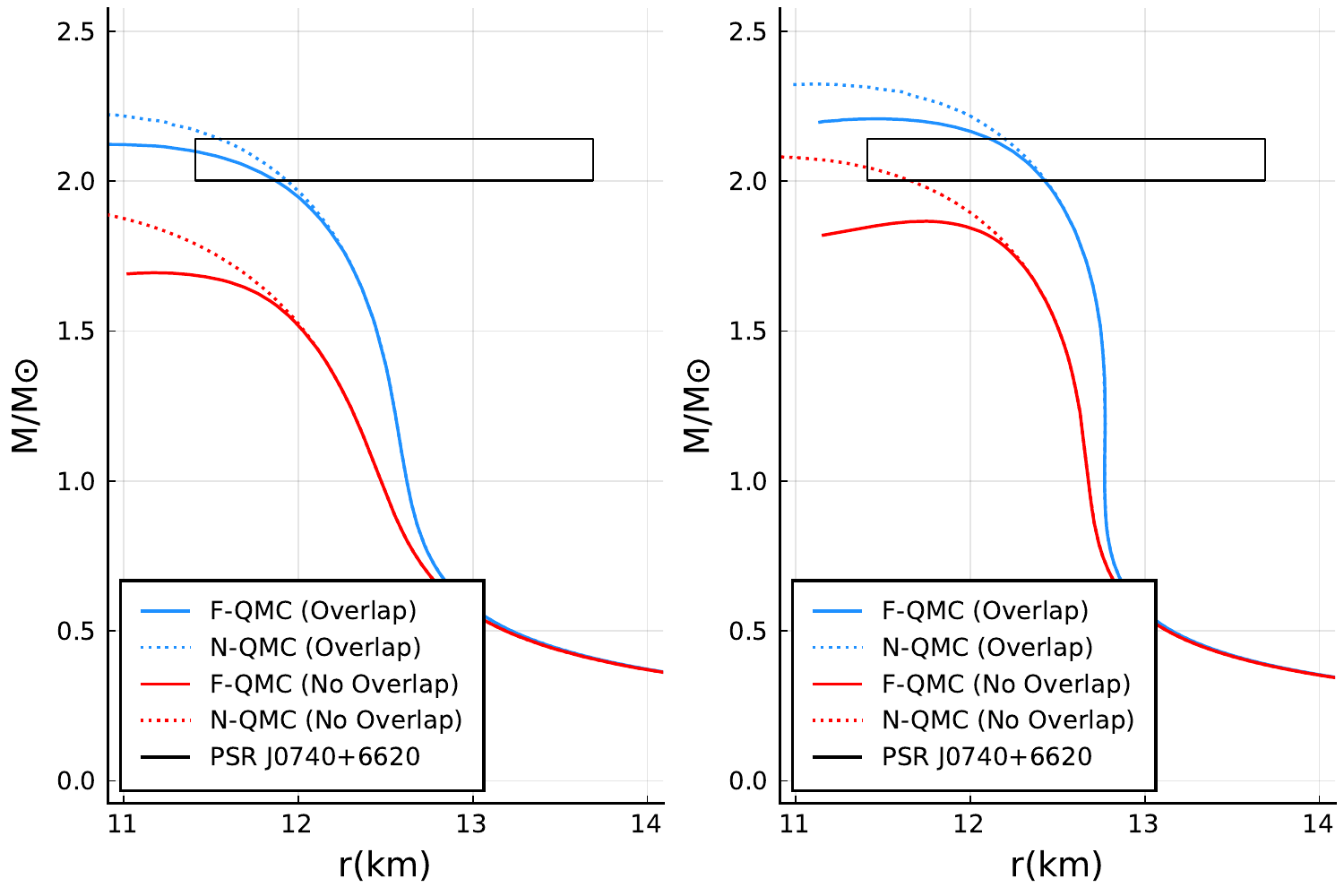}
    \caption{MR curve for $G_\delta=0$ fm$^2$. The left panel is $\lambda_3=0.02$ fm$^{-1}$, whilst the right panel is $\lambda_3=0.00$ fm$^{-1}$.}
    \label{fig:deltaMR}
\end{figure*}

\begin{figure*}
    \centering
    \includegraphics[scale=1]{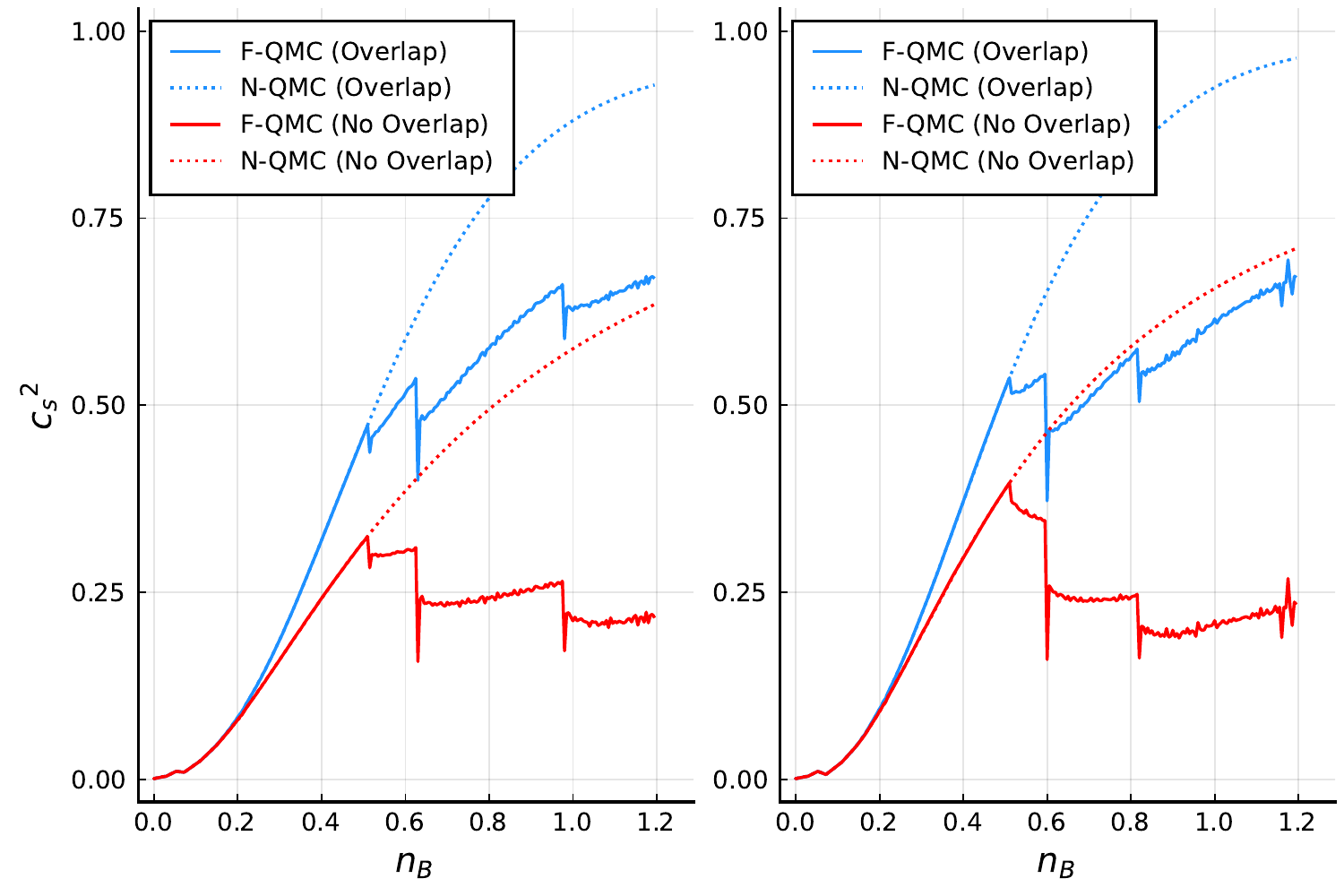}
    \caption{Speed of sound ($c_s$) when $G_\delta=0$ fm$^2$. The left panel corresponds to $\lambda_3=0.02$ fm$^{-1}$, whilst the right panel is $\lambda_3=0.00$ fm$^{-1}$.}
    \label{fig:deltaSound}
\end{figure*}

Figure~\ref{fig:deltaEoS} shows the EoS, with the corresponding MR curve in Fig.~\ref{fig:deltaMR}. As expected, there is a slight decrease in the mass and a substantial reduction in the radius. Without overlap and $\lambda_3$, the results show the same pattern reported in Ref.~\cite{Motta:2019tjc}. In Fig.~\ref{fig:Tidal} we saw that the $\lambda_3=0.00$ fm$^{-1}$ tidal deformation lay at the upper end of the experimental bound reported in Ref.~\cite{LIGOScientific:2018cki}. However, with the reduction of the radii resulting when the $\delta$ meson is excluded, QMC with overlap now falls within GW170817 constraints. Again the inclusion of the $\sigma^3$ term with $\lambda_3=0.02$ fm$^{-1}$ reduced the radius and fit the GW170817 constraint. 
Finally the speed of sound is shown in Fig.~\ref{fig:deltaSound}. The interpretation of these results may be found in section~\ref{sec:speed of sound}.

\bibliography{apssamp}% Produces the bibliography via BibTeX.
\begin{figure*} [h]
    \centering
    \includegraphics[scale=1]{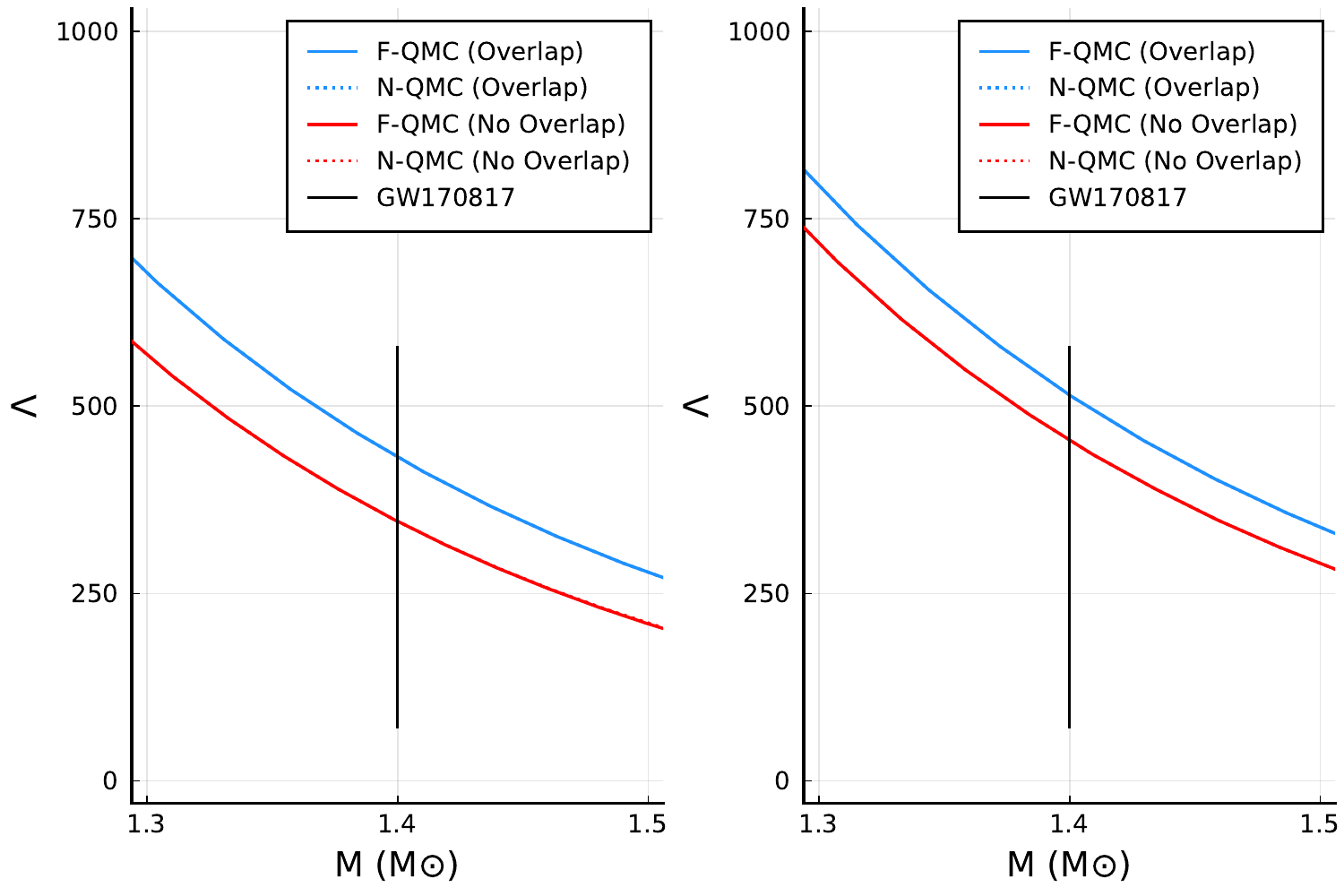}
    \caption{Tidal Deformation with $G_\delta=0$ fm$^2$. The left panel is $\lambda_3=0.02$ fm$^{-1}$, whilst the right panel is $\lambda_3=0.00$ fm$^{-1}$.}
    \label{fig:deltaTidal}
\end{figure*}

\end{document}